\documentclass[a4paper,twocolumn,11pt]{quantumarticle}

\pdfoutput=1
\usepackage[utf8]{inputenc}
\usepackage[T1]{fontenc}
\usepackage{amsmath,amssymb}
\usepackage{hyperref}

\newcommand{\ket}[1]{\left| #1 \right>} 

\begin{document}

\title[A farewell to waves]{Taking quantisation seriously: a farewell to waves}

\author{Geoff Beck}
\affiliation{Centre for Astrophysics and School of Physics, University of the Witwatersrand, 1 Jan Smuts Avenue, Johannesburg, WITS-2050, Gauteng, South Africa}
\email{geoffrey.beck@wits.ac.za}
\orcid{0000-0003-4916-4914}

\maketitle

\begin{abstract}
	The dual wave-particle nature of quantum objects is a notoriously unintuitive feature of quantum theories. However, it is often deemed essential, due to quantum objects exhibiting diffraction and interference. We extend the work of Land\'{e} and L\'{e}vy-Leblond to demonstrate that de Broglie wavelengths are not relativistically covariant as simultaneous spatial structures, making wave properties an unviable explanation of apparent interference. We then explore whether modern experiments vindicate an alternative view: that apparent waviness in diffraction and interference scenarios emerges as a consequence of quantised interactions between particles. Such a view has historically received very little attention, despite being the exact modern explanation of both the Kapitza-Dirac effect and ultrafast electron diffraction. We then study a photon orbital angular momentum realisation of the double slit to show that quantised exchanges can mimic interference. Finally, we demonstrate that the quantum formalism demands that a consistent momentum spectrum, determined at the point of scattering, undergirds double-slit interference patterns. This cannot be reconciled with the idea that particle momentum is determined by wave overlap beyond the slits.  
\end{abstract}


\section{Introduction}\label{sec1}

Feynman is often quoted as describing the double slit as containing ``the only mystery'' in quantum mechanics~\cite{feynman1977feynman}. This experiment is indeed curious, apparently requiring that otherwise material particles behave as waves some of the time. Such results inspired Bohr's complementarity: that neither particle nor wave is sufficient to encapsulate quantum entities alone. However, there are good reasons not to be sanguine about the physical reality of this waviness. First, we can note that these effects are apparent in the statistics of quantum experiments, with individual runs showing particle properties instead. As Popper has noted, in assigning wave nature to individual particles via their bulk statistics we commit the ecological fallacy~\cite{Popper1982-POPQTA}. This leads to the clumsy resolution of ``wavefunction collapse'', semi-dead cats~\cite{schrodinger}, and the interminable measurement problem~\cite{ballentine1998quantum,2007arXiv0712.0149W}. A further concern might be application of the Huygens-Fresnel principle to probability waves that cannot be disturbances in any sense. 

There is, however, a far graver problem with quantum waviness: the wavelengths are not covariant as correlation lengths in any form of relativity. In 1975 Land\'e argued~\cite{lande4} that because de Broglie wavelengths take the form $\lambda = \frac{h}{p}$ they must vary with inertial frame. Land\'{e} also assumed that all real waves should possess phase invariance, generating an apparent paradox. L\'{e}vy-leblond~\cite{leblonde} showed that $\lambda$ varying with inertial frame is necessary to maintain the Galilean invariance of Schr\"{o}dinger equation solutions and thus phase invariance does not hold for quantum waves. This establishes that $\lambda = \frac{h}{p}$ is not a Galilean invariant, but sheds no light on why this is the case. We demonstrate that $\lambda$ scales like the distance propagated by a particle, for both Galilean and Lorentzian cases. By this we mean that we cannot measure $\lambda$ as the simultaneous length of an extended structure in space. It thus cannot be invoked as a length scale to explain interference phenomena. 

Further evidence that real waviness is unnecessary can be found in experiments where the double slit is performed in single atoms~\cite{PhysRevLett.122.053204}. Here rubidium is photo-ionised by independent lasers of different frequency that, in combination, offer two possible ionisation paths. The resulting interference of photo-electron waves is understood to occur only in the abstract state-space, making it difficult to conclude that this interference requires actual waviness. The authors note that it is the quantised atomic transitions that provide phase coherence in the first place. This provides motivation to explore the role of quantisation as an explanation for phenomena that otherwise look like interference patterns formed by extant waves.

In particular, this work highlights a long-overlooked means to understanding apparent waviness in matter: quantised interactions. If incoming particles can only exchange discrete quanta of momentum with periodic scatterers (like slits or diffraction gratings), then we can describe ``interference'' patterns by local particle interactions following Duane's 1923 arguments~\cite{duane}. Here the grating is a quantum system, with allowed momentum states quantised according to its periodicity. Incoming particles interact with it and acquire quantised momentum along the same axis as the periodicity. Trajectories whose momenta match the quantisation rule arrive at ``constructive interference'' regions, while those requiring forbidden half-quanta yield ``destructive interference''. In fact, a very similar approach is used for explanation of the Kapitza-Dirac phase-grating effect~\cite{Batelaan_2000}. Here electrons are diffracted by a standing wave laser field. The diffraction peaks are sourced from absorption of quanta from the standing wave. 

Indeed, Land\'{e} proposed Duane's hypothesis as a solution to the de Broglie wavelength's lack of wave-like covariance~\cite{lande1965,lande4}. His arguments, however, amount merely to illustrating mathematical consistency, along with the allegation that his thesis is no more absurd than the quantum orthodoxy~\cite{lande4}. Once this is combined with the objection that local interactions cannot allow particles to ``experience'' an extended structure, like a diffraction grating, his arguments hardly seem compelling. This work aims to bring Duane's hypothesis into a new century with compelling evidence that was not available to earlier proponents. This demonstrates that serious, realist interpretations of quantum phenomena are possible, and require no extension of, nor additions to, the formalism. In this sense, we present an argument similar to that in \cite{kay2024escape}, but without association to Bohmian extensions of quantum mechanics. 

\subsection{Contribution to the literature}
Our work here provides novel arguments against the physical reality of quantum waviness. Both in terms of explaining the physics behind the de Broglie wavelength's lack of Galilean invariance, and in a derivation of the double slit momentum distribution at the point of scattering. We also bring together experiments with photon orbital angular momentum, ultrafast electron diffraction, coherence, momentum conservation, and the impact of polarisability on interference patterns to build an original case for Duane's model of quantum diffraction and interference. Additionally, a novel analysis of the quantum eraser highlights the power of this viewpoint in resolving mysterious phenomena and explaining why wave explanations arise in the first place. We believe this constitutes a new and unanswerable case against quantum interpretations that invoke physical wave properties, while simultaneously arguing for the reconsideration of a side-lined hypothesis.

\subsection{Paper structure}
This paper is structured as follows: in section~\ref{sec:debroglie} we demonstrate that de Broglie wavelengths are not associated with a simultaneously measurable length scale. Then, in section~\ref{sec:kd} we explore the description of the Kapitza-Dirac effect in terms of local particle interactions. In section~\ref{sec:duane} we explore how this particle description can be extended to other scenarios where diffraction and interference occur. To prove that quantised particle interactions are a viable alternative to interference, we explore experiments that support the position in section~\ref{sec:examples}. We next demonstrate that quantum mechanics requires the far-field double slit pattern is determined prior to wave overlap in section~\ref{sec:ds} and discuss ``which-way'' information and quantum erasers in section~\ref{sec:ww}. Finally, we draw conclusions in section~\ref{sec13}.  

\section{The problem of wavelengths}
\label{sec:debroglie}
As has been mentioned already, de Broglie wavelengths are not Galilean invariants. This is usually dismissed, as there exist covariant derivations of the de Broglie relation. However, it should cause us pause, a wavelength requires an extended structure, defined simultaneously at multiple points in space. If a de Broglie $\lambda$ is not a Galilean invariant, can it actually be such a length?  

Relativistic quantum theories use $p^\mu = \hbar k^\mu$ as a definition of a covariant wave-vector. This can be shown to lead directly to the de Broglie relation. Thus, we can provide a covariant mathematical description, but the wavelength is not a Galilean invariant. This can only mean that it does not merely scale with $1/\gamma$ under Lorentz boosting. The dependence on relative velocity must be different. This difference should allow us to determine what $\lambda$ actually is.

Thus, we consider the Lorentz momentum transform for a particle moving at $u$ in the same direction as the boost $v$
\begin{equation}
	p^\prime = \gamma\left(p - v\frac{E}{c^2}\right) = \gamma p \left(1-\frac{v}{u}\right) \; ,
\end{equation}
the de Broglie wavelength is then 
\begin{equation}
	\lambda^\prime = \frac{h}{\gamma p \left(1-\frac{v}{u}\right)} = \frac{\lambda}{\gamma} \frac{u}{u-v} \; .
\end{equation}
This is consistent with the Galilean limit when $\gamma \to 1$, the usual length transform is modified by the factor $\frac{u}{u -v}$. 

Now, the Lorentz transform for distances propagated by an object with speed $u$
\begin{equation}
	\Delta x^\prime = \gamma \Delta x \frac{u-v}{u} \; .
\end{equation} 
The structure is remarkably similar to the transform for the de Broglie wavelength. Thus, the lack of Galilean invariance arises because $\lambda$ is not an actual length of a structure that exists simultaneously in space (like a wave). The relativistic covariance matches that of a propagation distance instead, with no inherent simultaneity in any frame. Thus, we cannot interpret $\lambda$ as a length-scale which could explain phenomena like interference and diffraction without tacitly violating relativistic covariance. Furthermore, as there is no simultaneity to de Broglie waves, they are not waves. Actual wave phenomena do not suffer from this, their wavelengths are lengths and can be measured simultaneously. 

There is a snag, however. Electromagnetic waves obey the de Broglie relation and yet evidently have spatial extent, simultaneity, and measurable amplitudes. Does this demonstrate a failure of the argument above? No, as such a wave is not one, but a stupendous number of photons. Each photon is not a wave, but the above argument does not necessarily apply to their collective. Each may be thought of as a soliton-like pulse. A cascade of such pulses, correlated by their source, gives us what we need to build a real wave. Such wave-like coherence can indeed appear in the classical limit~\cite{bertet_complementarity_2001} and be analysed in terms of interactions, rather than ``waviness''~\cite{zeng2023}.

It is therefore not possible to ascribe actual waviness to quantum objects in a physically consistent manner, as is similarly concluded in \cite{Strnad_1985} (although our argument here is novel). All of this should cast some doubt on the physical, rather than mathematical, validity of de Broglie-Bohm theories, and potentially also on the correspondence between field modes and particles used in quantum field theory. A manifestation of the former may be seen in \cite{Pucci_Harris_Faria_Bush_2018}, where real pilot waves fail to reproduce the double slit experiment, possibly implying only those with physically inconsistent wavelengths can do so.

\section{The Kapitza-Dirac effect}\label{sec:kd}
This phenomenon, proposed in 1933~\cite{1933PCPS...29..297K} and later confirmed~\cite{1986PhRvL..56..827G,2001Natur.413..142F}, is where an electromagnetic standing wave acts as a diffraction grating for charged (or polarisable) particles. Ostensibly this represents a symmetry between light and matter, which is a reflection of the particle-wave duality. However, the KDE has a wave-particle object diffracted by a wave, whereas conventional diffraction has a wave diffracted by a classical barrier. This seems decidedly non-dual. The standard explanation is we are discussing a phase grating, i.e. spatially varying phase differences are applied to the matter wave by the standing light wave due to the electromagnetic coupling of the particles~\cite{Batelaan_2000}. We can immediately see that this is not actually explained by waves, both wave and particle properties are needed (unlike conventional optical diffraction).     

The presence of diffraction maxima in the KDE can be fully explained in terms of incident particles absorbing discrete light quanta of fixed energy~\cite{Batelaan_2000}. In particular, the particle (travelling initially in the $x$-direction) absorbs a photon with $y$-direction momentum $p_y = \hbar k$ ($k$ is the laser wavenumber) and subsequently emits a photon at $p_y = -\hbar k$ via stimulated emission from the reflected laser beam. Thus, the incident particle incurs a momentum change of $2\hbar k$. Since this can happen multiple times as the particle crosses the interaction region, we would expect a momentum spectrum for the out-going particles to be given in terms of $2 n \hbar k$, where $n$ is an integer. The intensity peaks are indeed found to correspond to these values.

We can explain the KDE completely in terms of particle interactions where the diffraction element is an active participant, we do not need to invoke any waviness. This is achieved by treating the diffraction element itself as a quantum system. Notably, we cannot actually describe the KDE without reference to particle properties, as even a wave picture must use particle interactions with the electric field to explain the wave's phase shifts~\cite{Batelaan_2000}. However, in barrier diffraction, a wave explanation treats the grating as a completely classical barrier to propagation and requires no reference to particle properties except at the point of detection. Perhaps we can symmetrise the KDE and other diffraction scenarios with a quantum description of the diffraction element? If possible, we would no longer require the physically inconsistent wave explanation. Furthermore, it would show that assigning a wave nature to quantum objects is contingent on regarding the diffraction element as classical.

\section{Diffraction by material gratings}\label{sec:duane}
We turn then to more prosaic diffraction phenomena: electromagnetic or matter waves diffracted by material gratings. These scenarios apparently \textit{demand} a wave-like description, even if this requires that a particle's wave aspect suffers from our subtle inconsistency with all forms of relativity. 

\subsection{Quantisation of momentum}
To facilitate our search for a picture where the diffraction element is no longer a passive, classical barrier, we will consider a 1D diffraction grating, centred at $x = 0$ and $y=0$, that extends symmetrically in the $y$ direction. A distant screen is used to detect diffracted particles at $x = L$, where $L \gg d$ (the grating periodicity scale). It is well known that the far-field diffraction maxima of a periodic grating can be found at angles $\theta$ (measured from the $x$-axis) given by 
\begin{equation}
	d\sin\theta = n \lambda \; ,
\end{equation} 
where $\lambda$ is the de Broglie wavelength, and $n \in \mathbb{Z}$. This can re-imagined via the de Broglie relation as 
\begin{equation}
	p\sin\theta = \frac{n h}{d} \; . \label{eq:deflect}
\end{equation}
Recognising that $p\sin\theta = p_y$, we can see that this is a quantisation rule on the transverse momentum of the outgoing particle. Particularly, interaction between incident particle and grating must involve the quantised exchange of momentum. Simply by reinterpreting a wave-interference formula, we have found an explicit discussion of the grating as participating in a quantum interaction. Our use of the de Broglie formula is merely heuristic, and we will see in section~\ref{sec:ds} that the momentum expression is the fundamental one extracted from the quantum formalism. Our interacting picture does not merely explain the diffraction maxima, as the minima can be similarly identified by
\begin{equation}
	p\sin\theta = \frac{m h}{2 d} \; , 
\end{equation}
where $m$ is an odd integer. 

Thus, the forbidden scattering angles, that apparently required a wave-based explanation, can be understood as being ruled out by requiring a forbidden half-quantum exchange of momentum. To elaborate, the $y$-momentum needed for deflection of the particle to ``destructive interference'' positions cannot be obtained from interactions with the periodic scatterer or its constituents, due to quantisation of collective momentum states (to be justified below). This quantisation rule was first observed by Duane~\cite{duane} in 1923, but has only received scattered mentions in the literature over the last century~\cite{breit1923,lande1965,VANVLIET196797,lande4,ballentine1998quantum,VANVLIET20101585,WENNERSTROM2014105,2018OptEn..57a5105M}. The generalised version of the simple analysis above fully reproduces both Laue and Ewald conditions~\cite{VANVLIET196797}. Notably, this viewpoint requires that the diffraction grating must be characterised as inherently quantum, removing the artificial separation of quantum and classical within the experiment. 

\subsection{Bloch's theorem and the M\"{o}ssbauer effect}

Duane's model does come with a difficulty: how to justify the quantised nature of the interaction? Mathematical justification is easy to find~\cite{VANVLIET20101585,WENNERSTROM2014105} in Bloch's theorem~\cite{bloch1929}, which requires that structures characterised by spatially periodic potentials have quantised momentum eigenstates. For a crystalline solid, these are particular collective motion states or phonons. The scattering interaction could therefore be similar to the physics surrounding the M\"{o}ssbauer effect~\cite{mossbauer}, where the incident particle interacts with a local lattice site and recoil can be absorbed by collective motion~\cite{mossbauer2,mossbauer3}. The interactions that mediate collective motions naturally restrict local recoils as well, hence the points of ``destructive interference'' corresponding to forbidden lattice momenta. Thus, the intensity peaks correspond to resonant scattering, where exact lattice quanta are exchanged, and the continuity is supplied by local recoils or a low energy background of excited phonon states. 

Notably, the explanation of the M\"{o}ssbauer effect eliminates the problem that stymied Land\'{e}~\cite{lande4}. It is possible for a local interaction to depend upon the global structure of a participant. Importantly, this explanation hinges upon the periodic structure of the diffraction grating, with Bloch's theorem assuring us it will extend to any periodic diffraction phenomenon and \cite{VANVLIET20101585} extending it to less symmetric structures. However, there is no reason that the same physical mechanism should explain every example of such interference. Symmetry guarantees interference mathematically, but does not uniquely specify any mediating physics.

\subsection{Does this conflict with the uncertainty principle?}

The looming question is whether such a model can be compatible with the uncertainty principle, as it is commonly held that a particle cannot possess definite momentum and position at the same time (see e.g. \cite{tamvakis2019,townsend2012modern,mcintyre2012quantum}). This is not a problem for Duane's model, as the aforementioned belief is a fallacious use of the principle~\cite{ballentine1998quantum,Popper1982-POPQTA} with several popular texts acknowledging this~\cite{rae2002,Sakurai:1167961,Griffiths2004Introduction,moore1998}. As rigorously defined, the uncertainty principle applies to standard deviations over a measurement ensemble~\cite{kennard_zur_1927}, it thus has nothing to say about the ontic status of individual particle properties. Indeed, Popper argued that were the principle to apply to individual particles, the minimum ensemble deviation could never be probed~\cite{Popper1982-POPQTA}. 

\section{Evidence supporting Duane's model}
\label{sec:examples}
Interestingly, there are experiments that already exhibit a hallmark of Duane's mechanism as well as effects that undermine wave explanations. These will be detailed below.

\subsection{Bragg diffraction}
When X-rays are incident on a crystal lattice they are well-known to experience preferential reflection to discrete angles. Interestingly, electrons at the same energies experience a somewhat different behaviour. Allegedly, both of these particles correspond to the waves of the same wavelength. However, despite experiencing the same boundary conditions, these waves experience different diffraction/interference behaviours~\cite{cowley-diffraction}. Particularly, the electrons produce many more diffraction spots than X-rays, and these spots are approximately patterned after the reciprocal lattice of the scatterer. The explanation is simple: electrons interact far more strongly with the crystal than photons do. However, this already betrays the wave explanation. It is the particle interactions that determine the resultant pattern, and the quantisation of diffraction angles is determined by the reciprocal lattice vector. This is precisely what is implied by the more general form of Eq.~(\ref{eq:deflect}), as demonstrated in \cite{VANVLIET196797} which also derives the Laue and Ewald conditions. In addition, the reciprocal lattice projection is the momentum representation of the crystal's periodicity and only occurs perpendicular to the electron beam direction for transmission-type experiments. The electrons are thus gaining quantised momenta, determined in both magnitude and direction by the lattice periodicity.  

\subsection{Ultrafast electron diffraction}
One expectation of Duane's model would be sensitivity of diffraction patterns to excitation of phonon modes within the diffraction element prior to scattering. The presence of excited phonon modes, that might be negligible in ordinary conditions for the target, will change the possible interactions with the scatterer. This being consistent with the assertion that all diffraction is a consequence of interactions with the incident particles. One might be tempted to invoke a wave and particle explanation here, waves result from boundary conditions and the phonon interactions contribute a distinct extra component. However, this is untenable given existing experiments~\cite{electron-phonon1,electron-phonon2,photon-phonon}. 

In \cite{electron-phonon2} it is demonstrated that exciting phonons alters the diffraction spectrum in a very particular manner. The apparent effect is that phonon excitation shifts intensity away from the Bragg peaks to a surrounding continuum. The more rapidly the probe follows the excitation, the more pronounced this effect is. Thus, there are not two separate effects: wave interference and phonon interactions. In fact, the phonon interaction appears to suppress the distinct interference peaks~\cite{electron-phonon2}. This makes sense if we are discussing competing interactions, something strongly suggested by the form of the wavenumber being exchanged between particle and scatterer $Q = q + K_Q$~\cite{electron-phonon1}, where $q$ is the phonon wavenumber and $K_Q$ is the closest reciprocal lattice vector to $Q$. This is very clearly a momentum quantisation condition. When no phonons are excited, the exchange is quantised according to the reciprocal lattice vector, which follows from the general form of Eq.~(\ref{eq:deflect})~\cite{VANVLIET196797,VANVLIET20101585}. When a significant phonon background is present, they supply the additional momentum $q$ to disperse the strict Bragg quantisation. Notably, the final pattern depends upon both the phonon energies and their relative population. These phonons need not be viewed as waves, as they represent collective lattice motions realised by local sites.

Additionally, the phonon spectrum prior to excitation also plays a role~\cite{electron-phonon2}. Given that phonon modes distributed intensity away from the quantised Bragg peaks, this demonstrates is a case where continuity is supplied between otherwise discrete peaks, even in an unexcited target. We thus have a possible mechanism answering precisely the question we should ask about Duane's mechanism: why is the spectrum often continuous if the momentum states are strictly quantised? While it is unlikely to be the same mechanism that answers the question in general, it does suggest that physical processes exist that perform the required role.

\subsection{Diffraction of polarisable molecules}
Experiments with polarisable molecules demonstrate that van der Waal's interactions during diffraction contribute to the pattern, in a manner degenerate with the slit width~\cite{grisenti_determination_1999} (similar phenomena are used to argue against wave-particle duality in \cite{Passon2022}). In these experiments, beams of molecules are diffracted by silicon nitride gratings. The diffraction patterns are found to depend on the dipole polarisability of the incident molecules. As previously mentioned, the effect of the extra interactions produces a modified ``effective slit width''. This is very suggestive of Duane's explanation, which connects slit width to interactions already. It should be noted that further evidence of interactions contributing to the diffraction pattern can be found in \cite{Batelaan_2016}.

Interestingly, these additional interactions also lift the zeroes of the diffraction pattern. This is unsurprising if diffraction momentum changes are being supplied by interactions in general, as the van der Waal interactions can allow deflection angles otherwise forbidden by the intrinsic lattice momenta. More crucially, this case supplies direct evidence that local interactions within diffraction gratings can supply an effect that is degenerate with the boundary conditions of apparent wave interference, precisely what is being claimed by Duane. 


\subsection{Coherence}
Diffraction experiments like the double slit are often stated to require ``spatial coherence'' of incoming particles. However, this is severely questionable, as \cite{bernstein_measurement_1987} shows that only collimation is needed. Indeed, coherent and incoherent waves are indistinguishable, provided they have negligible dispersion in perpendicular momentum on the scales relevant to the experiment. This already seems to be both powerful evidence for the particle picture presented here and against a wave explanation.

The collimation requirement makes complete sense for a particle picture. Double slit interference requires an incident particle distribution whose momentum statistics will not obscure the effect of the double-slit scattering itself. If one wishes to preserve a wave view, the only explanation can be that coherence is imposed on incoherent incoming waves by their interaction with the slits. This is very difficult to square with the slits behaving as a classical, inert barrier. Clearly, the interaction requirement is very consistent with the notion that wave explanations emerge from ignoring interactions, as is similarly argued in \cite{zeng2023}.   

More generally, a multimode photon source has a coherence length because differences in photon momenta become apparent over a long enough distance, hence the dependence of a coherence length on bandwidth ($\delta p_\parallel$ in the direction of motion) and the speed of light. In a diffraction scenario this would produce blurring via incompletely overlapping patterns for each mode, as their deflection angles will all differ due to dependence on $\frac{p_\perp}{p_\parallel}$. 

Another interesting aspect of coherence is that longitudinal coherence lengths do not change as a quantum wave packet spreads~\cite{klein_longitudinal_1983}. This can lead to situations where waves overlap but do not interfere, an extremely counterintuitive, yet experimentally confirmed~\cite{kaiser_direct_1983}, result for wave explanations of quantum phenomena. In some scenarios the coherence length and size of the wave packet can even become completely unrelated~\cite{jabs_comment_1987}. This teeters on the brink of total contradiction.



\subsection{Momentum conservation}

It is commonly asserted that conservation laws in quantum mechanics work only at the statistical level. This being due to the fact that the formalism can only provide such rules for the averages of conserved observables. A particular issue is that of superpositions. If a superposition collapses then there is some non-conservation occurring. 

A particle viewpoint, as suggested by Duane, requires non-statistical conservation of both momentum and energy. Importantly, experimental observations of recoiling double-slit analogues~\cite{liu_einsteinbohr_2015} require an assumption of momentum conservation on the individual particle level~\cite{Batelaan_2016}, in order for the conventional explanation to succeed. Additionally, single particle orbital angular momentum conservation was observed in \cite{kopf_conservation_2025}. Finally, there exists a theoretical justification to generalise this in \cite{aharonov_conservation_nodate,collins_conservation_2025}. 

If momentum is indeed individually conserved in double slit scenarios, it is miraculous for wave explanations. It requires that the slit recoil is determined retrocausally. That is, after the particle's momentum is determined by capture at the screen. However, such conservation is perfectly natural in Duane's picture.

\subsection{The angular double slit}\label{sec:oam}
Duane's picture is not limited to diffraction gratings. Indeed, it was quickly shown to apply to the double slit as well~\cite{breit1923}. Modern experiments can offer us an analogy to demonstrate that quantised exchanges can produce ``interference patterns''. To do this, we consider a double slit experiment where the momentum analogue is inherently quantised. Thus, it can only be changed by discrete quanta. If we can obtain the analogous phenomena to a regular double slit, we have unambiguous, empirical evidence of the viability of Duane's mechanism. 

To this end, we can construct an analogue to the double slit using the Orbital Angular Momentum (OAM) of light, $\ell$, as is done in \cite{Jack_2008}. In the angular double slit, a spatial light modulator displays two angular sections of a fork hologram. The net result is the incident beam receives OAM from both the angular slits and the fork hologram (the latter allowing for spectral decomposition of the outgoing beam). The spectral prediction is given by~\cite{Jack_2008}
\begin{equation}
	P(\ell) \propto \mathrm{sinc}^2\left[\frac{\beta\ell}{2}\right] \cos^2\left[\frac{\alpha \ell}{2}\right] \; , \label{eq:prob-l}
\end{equation}
where $\alpha$ is the slit spacing and $\beta$ their width. This is verified empirically~\cite{Jack_2008} and can be compared to the linear form
\begin{equation}
	P(p_y) \propto \mathrm{sinc}^2\left[\frac{\sigma p_y}{2\hbar}\right] \cos^2\left[\frac{\Delta_y p_y}{2\hbar} \right]\; , \label{eq:prob-p}
\end{equation} 
where $p_y$ is the transverse momentum, $\sigma$ is the slit width, and $\Delta_y$ is the spacing. Writing $\alpha = f\pi$, a simple correspondence can be established between the two by introducing the unit-free momentum $\tilde{p} = 2\frac{\Delta_y p_y}{f h}$ so that
\begin{equation}
	P(\tilde{p}) \propto \mathrm{sinc}^2\left[\frac{f \pi \tilde{p}}{2}\frac{\sigma}{\Delta_y}\right] \cos^2\left[\frac{f \pi \tilde{p}}{2} \right]\; . \label{eq:prob-p-tilde}
\end{equation}
The factor of 2 in $\tilde{p}$ corresponds to the 2-fold rotational symmetry of the angular mask~\cite{Jack_2008}. 

A vital aspect of this experiment is that $\ell$ modes are ``free-space modes''~\cite{Sroor_2021}. Meaning a beam does not change $\ell$ while propagating in free space. Thus, a beam's angular momentum is set by a discrete exchange at the spatial light modulator and not determined at some point of overlap at the detector. This is consistent with the ability to perform the experiment with classical beams. So, our $\ell$ spectrum can be simply interpreted as the fraction of beam power propagating in mode $\ell$. 

\begin{figure}[ht!]
	\centering
	\resizebox{0.8\hsize}{!}{\includegraphics{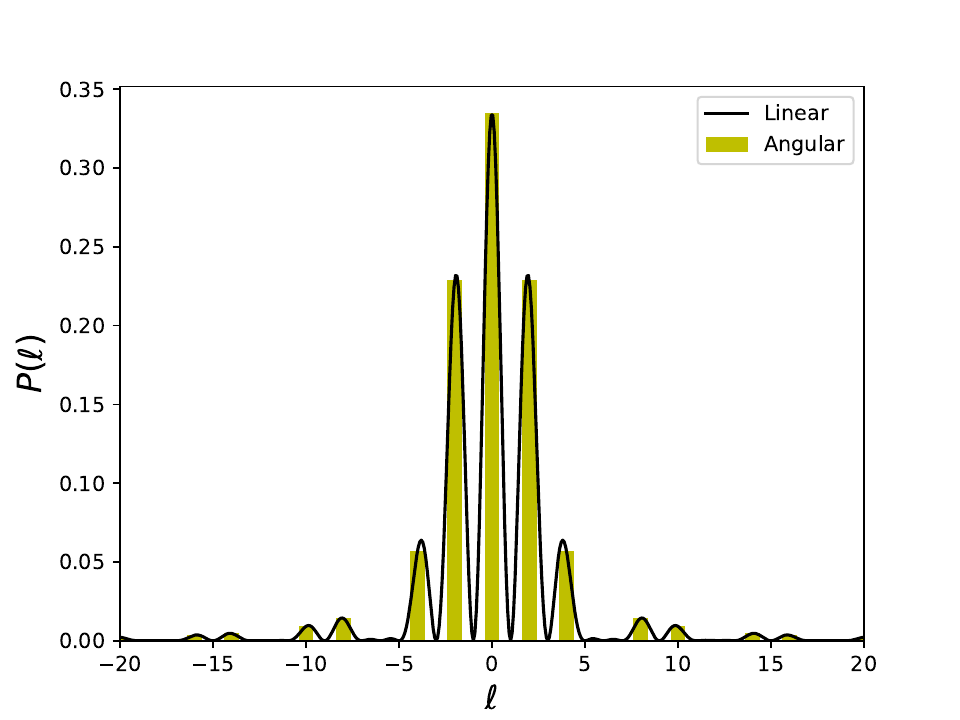}}
	\caption{OAM spectrum from angular double slit when $\alpha = \pi$, $\beta = \frac{\pi}{3}$ plotted as a bar graph. The line plot shows a linear double-slit momentum distribution in $\tilde{p}$ when $\frac{\Delta_y}{\sigma} = 3$ and $f = 1$.}
	\label{fig:oam}
\end{figure}

We display an example OAM spectrum from such an arrangement in Figure~\ref{fig:oam}. This pattern can be simply understood as resulting from an interaction between the slits and beam that imparts discrete units of $\ell$. It is notable that our $\ell$ spectrum, although discrete, is perfectly matched by the linear double slit far-field intensity pattern (or momentum representation distribution). This analogy provides evidence that a discrete exchange interaction can, in empirical fact, produce something that looks like an interference pattern. Since $p$ is not inherently quantised, but reproduces the case where $\ell$ is, we have to conclude that quantisation is being imposed upon $p_y$ via interaction with the slits.

\section{The double slit revisited}\label{sec:ds}
An intriguing aspect of the OAM double slit in section~\ref{sec:oam} is that it does not matter (in principle) where we measure the OAM spectrum. Both near-field and far-field diffraction regimes yield the same spectrum. For wave-based explanations this result is extraordinary. The ``interference'' pattern we see in the far-field is already imprinted in $\ell$ (or by analogy $p_y$) within the near-field. It is easy to demonstrate that this is so. Consider a position wavefunction at the double slit position $\psi(\vec{x})$. This can be written in the momentum representation in terms of Fourier coefficients $a_k$ (for each momentum $k$). The momentum spectrum is then $\vert a_k\vert^2$. In this same plane wave basis, propagation in free space merely accumulates a phase factor, or $a_k e^{i \vec{k}\cdot\vec{x}} \to a_k e^{i \vec{k}\cdot\vec{x}} e^{i\vec{k}\cdot\vec{\delta x}}$ after propagating $\vec{\delta x}$ from $\vec{x}$. Thus, the Fourier coefficients, and hence the $p$ spectrum, are unchanged by free-space propagation. The same argument holds for the angular case when using, for instance, the Laguerre-Gauss basis~\cite{Sroor_2021}. The fact that changes occur in the position space pattern, but not the momentum spectrum is then down to each mode accumulating differing phases (Gouy phase). We will revisit this from a particle point of view later.

What the above argument demonstrates is that a consistent momentum distribution undergirds interference patterns, rather than requiring particle momenta be determined by wave-overlap at the point of detection. 
In fact, the Gouy phase also demonstrates each quantum object has a consistent momentum, i.e. it cannot possess a history without a consistent trajectory. Otherwise, it would accumulate a mixed Gouy phase by oscillating in $p$ during propagation. 
To make this complete, we can demonstrate that the quantum formalism itself requires that the momentum spectrum is determined before wave over-lap can even occur. 

\subsection{Double slit momentum distributions}
We start by formulating a simple, position-representation amplitude for particles passing in the $x$-direction through the linear slit apparatus. This is positioned at $x=0$ and extends symmetrically in the $y$-direction. Thus, we use the normalisable state
\begin{equation}
	\psi(y) \propto \Pi\left(\frac{y-\Delta_y/2}{\sigma}\right) + \Pi\left(\frac{y+\Delta_y/2}{\sigma}\right) \; ,
\end{equation}
where $\Pi\left(\frac{y}{a}\right)$ is the rectangular function of width $a$, centred at $y = 0$. It should be noted that there are no interference terms contributing to $\vert \psi\vert^2$, which just corresponds to an unremarkable distribution for particles occupying any point within either of the two slits with equal probability. This $\psi(y)$ is valid only at the slits' position ($x=0$) on the propagation axis. The momentum amplitude is then found via the usual Fourier transform
\begin{equation}
	\phi(p_y) \propto \int_{-\infty}^{\infty} \psi(y) e^{\left(-i \frac{ \vec{x}\cdot\vec{p}}{\hbar}\right)} dy \; .
\end{equation}
This results in
\begin{equation}
	\phi(p_y) \propto \mathrm{sinc}\left[\frac{\sigma p_y}{2\hbar}\right]\left(e^{i\frac{\Delta_y p_y}{2\hbar}} + e^{-i\frac{\Delta_y p_y}{2\hbar}} \right) \; ,
\end{equation}
where any contributions from orthogonal momenta amount merely to global phase factors. The Fourier integration over all $y$ can be understood here as considering the global structure of the scatterer, rather than invoking interfering plane waves.
Our momentum distribution is then
\begin{equation}
	\vert \phi(p_y) \vert^2 \propto \mathrm{sinc}^2\left[\frac{\sigma p_y}{2\hbar}\right] \cos^2\left[\frac{\Delta_y p_y}{2\hbar} \right]\; . \label{eq:p-deriv}
\end{equation} 
This indicates that the $p_y$ spectrum, with both single and double-slit factors, is already imposed at $x=0$, before waves originating from each slit get a chance to overlap. 

Thus, actual spatial wave-overlap and interference is irrelevant in forming the far-field position distribution, which precisely mirrors $P(p_y)$ above. This conforms nicely with the $\ell$ analogy, as the beams in question do not change $\ell$ while propagating in free space. Meaning their $\ell$ must only be determined at the spatial light modulator, not in some interference region at the detection point (unless one wishes to argue the detector itself couples different modes based on its distance to the slits). 

Moreover, the maxima of $\cos^2\left[\frac{\Delta_y p_y}{2\hbar} \right]$ occur when $p_y = \frac{n h}{\Delta_y}$, the same quantisation rule observed by Duane, as $\Delta_y$ is the symmetry scale of the slits. Although this is a limited translation symmetry, rather than a full periodicity, momentum quantisation appears following similar arguments~\cite{VANVLIET20101585}. Additionally, the factor $\mathrm{sinc}^2\left[\frac{\sigma p_y}{2\hbar}\right]$ has maxima where $p_y = \frac{m h}{2 \sigma}$ where $m$ is an odd integer. These maxima are spaced out by integer values of $\frac{h}{\sigma}$ and the minima occur $\frac{h}{2\sigma}$ from the maxima. Since $\sigma$ represents a second symmetry scale for reflection around the slit centre, the presence of two simultaneous quantisation rules is unsurprising. The ``missing orders'' occur when minima $\frac{m^\prime h}{\sigma}$, $m^\prime\in \mathbb{Z}$ match maxima $\frac{n h}{\Delta_y}$. If we let $\sigma = f\Delta_y$ this matching condition means when $\frac{m^\prime}{f}$ is an integer we find a ``missing order'' that was disallowed by the $\sigma$ scale quantisation. 

It is clear then that the interference-like behaviour in the diffraction of quantum systems can be consistently viewed as originating from the quantisation of particle momenta. This is evidenced directly by the angular double slit, as well as other experiments invoked in section~\ref{sec:examples}. Additionally, the momentum spectrum is determined at the slits and then remains unchanged during propagation (as shown by the far-field pattern reproducing it) in complete harmony with Duane's hypothesis. This holds even when considered within the mathematics usually attributed to the wave nature of quantum objects.

\subsection{Gouy phase}
What then is the particle equivalent of the mode-dependent phase accumulated during propagation? For waves, this phase determines the variation of the position-space image with distance between the slits and screen. The obvious answer is to consider whether the particle's position in the slits affects its destination. This would be apparent in the near-field, but by definition unnoticeable in the far. Thus, we could simply make geometric projections for particles originating from different points in the slits, each sampled from the same momentum distribution above. 

However, we will not reclaim the full detail of a Fresnel pattern. This is because the derivation above does not expose the details of interactions between incident particles and slits~\cite{2018OptEn..57a5105M}. In that, there is no explicit in-slit position dependence. We should expect particles passing closer to the edges have differing interaction probability with each edge and will contribute to the spectrum differently to a particle passing exactly between both edges. Our above result is the correct total spectrum (integrated over the slit width), but lacks the detail that would result in something of similar complexity to a Fresnel integral for the near-field diffraction pattern. 

We thus suggest the different propagation phase for each wave mode is tracing their in-slit position associations. This also provides a very natural explanation of why diffraction images converge to a far-field form, as this is the long-distance geometric projection of trajectories sampled from the total momentum spectrum.

\section{``Which-way'' information}\label{sec:ww}
Now we turn to ``which-way'' information. This is a phenomenon where determining which slit a scattered particle traversed can remove the interference pattern~\cite{feynman1977feynman}. More interestingly, an uncertain determination of the path merely reduces the contrast of the pattern in proportion to the uncertainty~\cite{greenberger_simultaneous_1988}. This appears to be strong argument for wave behaviour, as by `marking' the waves emanating from different slits, we are removing the essential coherence needed to produce interference. However, the incoming wave does not need any actual wave coherence~\cite{bernstein_measurement_1987}, so this explanation fails. 

We can address this quite simply in ``particle language'' instead. To illustrate, we will consider the case of determining ``which path'' information by scattering photons off of particles as they emerge from the slits. We choose this case for physical transparency. It is well known that the spatial resolution of such scattering detections is limited by probe wavelength. However, this can be easily proven for a pure particle approach. All we need do is regard a particle emerging from the slits and absorbing a probe photon as being excited. This excitation has an average lifetime $\tau$ that obeys $E \cdot \tau \geq \frac{\hbar}{2}$ where $E$ is the excitation energy. Thus, the position uncertainty of this probe is the distance moved as a result of absorption of the probe photon prior to re-emission, or $\Delta x = \Delta v \tau$. Here $\Delta v$ is the velocity kick introduced by absorption. Now we can write
\begin{equation}
	\Delta x \geq \Delta v \frac{\hbar}{2 E} = \frac{\Delta v \hbar}{m \Delta v^2} \; .
\end{equation}     
Where the second equality assumed that we can ignore the initial trajectory of the particle (we are only interested in the excitation and can happily choose the initial rest-frame of the particle). Since $m\Delta v = p_\gamma$, or the probe momentum, we have that $\Delta x \geq \frac{\hbar}{p_\gamma}$. Now we can see that discriminating between slits will require photons with momenta $\sim \frac{h}{\Delta_y}$, very similar to the momentum quantisation responsible for the apparent interference pattern. Moreover, we can imagine that the particle's $p_y$ is randomly changed during the probe scattering, as there are no preferential scattering angles. We can implement this by convolving $P (p_y)$ with a flat window-function between $p_y - \frac{h}{\alpha \Delta_y}$ and $p_y + \frac{h}{\alpha \Delta_y}$ where $\alpha \in \mathbb{R}$. Higher resolution, i.e. $\alpha < 1$, results in a broader convolution that reduces the contrast of the bright and dark fringes.

\subsection{Polarised double slits}
What about the more experimentally practical ``which-way'' case? Where photons pass through a double slit where each slit has one of a pair of orthogonal polarisers before or after it. Here the physics is far less transparent. However, we can see that such a situation disrupts the symmetry of our diffraction set up. Thus, we should expect it to alter the outgoing photon momentum. Specifically, we should see some modification to the double slit quantisation effect but not that of the single slits. In the case where photons can only traverse one slit, as a result of the polarisers, we should lose the double slit pattern entirely. This is what gives rise to the apparent need for wave overlap. However, we can see it is a mathematical consequence of lost symmetry. This does nothing to identify physical processes at work, as we have no detail of the interaction between photon and scatterer. However, in the case of modified, rather than destroyed, symmetry we can make more progress. 

Consider a double slit being illuminated by unpolarised light. One slit is preceded by a right-handed ($R$) circular polariser, and other left-handed ($L$). This famously yields ``no interference'', or, more accurately, only a single-slit diffraction pattern. However, if we interpose a linear polariser before the screen, a full interference pattern magically returns. In fact, there are \textit{two} patterns, one for each orthogonal linear polarisation~\cite{PhysRevA.65.033818} (and a range of intermediate combinations for different axis choices). One such pattern has the same form as the conventional double slit. The other has a central point of destructive interference, rather than a maximum. This is a baffling experiment if one is committed to an explanation in terms of interfering waves. The orthogonal, circular polarisations surely prevent meaningful wave overlap, yet a linear polariser restores an interference pattern. If the linear polariser changes our waves to allow overlap again, why does the propagation distance from the slits (not linear polariser) determine the observed pattern? This has lead to somewhat fanciful interpretations in terms of ``which-way'' information, which is erased by the linear polariser. Furthermore, we cannot invoke ``coherence'' to provide a more objective wave explanation~\cite{bernstein_measurement_1987} without admitting that it is interactions that shape this ``interference'', rather than waviness.

This remarkable result is far easier to understand in terms of particle interactions. Our two polarisers have opposite handedness, so we can consider an opposite parity double slit position amplitude
\begin{equation}
	\psi(y) \propto \Pi\left(\frac{y-\Delta_y/2}{\sigma}\right) - \Pi\left(\frac{y+\Delta_y/2}{\sigma}\right) \; .
\end{equation}
It can be easily shown that a Fourier transform yields
\begin{equation}
	\vert \phi(p_y) \vert^2 \propto \mathrm{sinc}^2\left[\frac{\sigma p_y}{2\hbar}\right] \sin^2\left[\frac{\Delta_y p_y}{2\hbar} \right]\; . \label{eq:p-deriv-odd}
\end{equation} 
In other words, an odd-parity amplitude leads to a momentum spectrum with a central $0$. This trivially adds with the even version to produce $\mathrm{sinc}^2\left[\frac{\sigma p_y}{2\hbar}\right]$, a pure single slit pattern. Surprisingly, these odd and even states form a single-slit momentum pattern under either amplitude or probability addition. It must be emphasised that the source of any parity differences must be traced to an interaction between photons and the polarisers.

We can now understand our ``magic eraser''. \textit{Complementary interference patterns are simultaneously visible without the linear polariser}. One of a complementary pair is revealed by the linear polariser, regardless of axis choice. This is a critical signature of an interaction explanation, as these interactions are not turned off by the existence of ``which-way'' information. 

The source of the opposite parity patterns can be found if we decompose our circular polarisations into
\begin{align}
\ket{R} & = \frac{1}{\sqrt{2}}\left(\ket{H} - i\ket{V}\right) \; , \\
\ket{L} & = \frac{1}{\sqrt{2}}\left(\ket{H} + i\ket{V}\right) \; . 
\end{align}
Thus, we must conclude that the $\ket{H}$ and $\ket{V}$ states (or any similar orthogonal pair) carry the two different interference patterns seen with this choice of basis for the final polariser. Moreover, we also know each slit emits both interference patterns. These two states having opposite parity in the $\ket{R}$, $\ket{L}$ basis completes the link. 

Our conclusion is that this is still susceptible to our interaction analysis, despite the failure of a wavey coherence attempt. We have modified the global symmetry via the orthogonal polarisers and thus altered the final momentum distribution. The operation of the polarisers alters the photon states, creating correlations between polarisation and momentum, as implied by sharing the same mathematics as a Stern-Gerlach apparatus, although this seems more subtle for photons. These correlations establish the parity difference that translates to the final momentum distribution. Crucially, we see the symmetry is an aspect of both the incoming photons and the scattering slits, this is not revealed until we impose this parity difference on the photons entering the slits. The magic of quantum mechanics is that we can predict the correct outcomes from symmetry principles, even without a characterisation of the particle interactions. This is not mysterious in a particle view, as the outcome of the interactions is constrained by quantisation, which is imposed by the symmetry.


\section{Conclusion}\label{sec13}
The de Broglie wavelength is not covariant as the length scale of a simultaneous extended structure in any reference frame. Thus, we cannot use it as physical explanation of observed phenomena, it is merely a useful heuristic. This puts claims about wave-particle duality under extreme strain and undermines otherwise compelling explanations of interference and diffraction. 

To resolve this, we have argued that the apparent wave nature of quantum objects is only manifest when diffraction elements are treated classically~\cite{zeng2023}. In contrast to the Kapitza-Dirac effect, which never needs to invoke either the waviness of incident electrons or the scattering photons. When the diffraction element is considered as a participating quantum system, a picture of diffraction and interference emerges that describes them as apparent consequences of quantised momentum exchanges. Thus, uniting phase-grating and barrier diffraction experiments with one physically consistent explanation. This is justified by the periodic structure of the scatterer implying it must have quantised momentum eigenstates. 

We have also presented experiments that signal the validity of our explanation of waviness: ultrafast electron diffraction, where phonon excitation alters diffraction patterns, and polarisable molecule diffraction, with a degeneracy between slit width and interaction strength. Further points of interest were the lack of a coherence requirement in diffraction experiments, in favour of collimation instead~\cite{bernstein_measurement_1987}, and indications of momentum conservation on a per event level~\cite{Batelaan_2016,kopf_conservation_2025,collins_conservation_2025}. Moreover, a double slit analogue with the explicitly quantised photon orbital angular momentum shows the same expected distribution as the linear double slit momentum. This provides evidence that a quantised exchange interaction can produce what looks like an interference pattern. Further, the only conclusion to be drawn from the analogy is that the interaction is responsible for quantising linear momentum in a regular double slit. 

We revisited the double slit and demonstrated the momentum distribution is determined at the slits themselves, not at the point of particle detection and before wave overlap could even occur. In addition, it does not change during propagation to the screen. Combined with momentum conservation~\cite{Batelaan_2016} and issues of coherence~\cite{bernstein_measurement_1987}, this is a powerful argument in favour of discarding wave overlap as a physical explanation. Finally, we broadened this discussion to include ``which-way'' information. Showing that this can even be shorn of mystical associations in the process. In particular, we analysed a ``quantum eraser'' to demonstrate that interference patterns always exist in this scenario, they simply occur in complementary pairs as a consequence of parity symmetry being broken by polarisers. 

This work demonstrates that complex amplitude interference is capable of representing particle physics that it does not resemble at all. Thus, explaining why naive positivism resulted in a view of quantum mechanics at odds with realism and logic. Interestingly, the source of both the quantisation and apparent waviness is symmetry, explaining why we can employ an inconsistent physical picture but obtain the correct predictions anyway. This link also explains why group theory has such utility in quantum mechanics: the tools of symmetry are naturally useful to describe quantised systems where the quantisation itself arises from symmetries.

While we do not cover interference with beam splitters, arguments have already been provided that these cases must be understood in terms of interactions between particle and beam splitter that involve exchange of energy and momentum~\cite{zeng2023}. This is entirely in keeping with the main thesis here, that we should not regard apparatus as inert in quantum experiments.

\section*{Acknowledgements}

Thanks to Andrew Forbes, Thomas Konrad, Robert de Mello Koch, and Sam van Leuven for helpful discussions; as well as Emil Roduner and Tjaart Kruger for their encouragement.

\bibliographystyle{iopart-num}
\bibliography{waves}

\providecommand{\newblock}{}
\begin{thebibliography}{10}
\expandafter\ifx\csname url\endcsname\relax
  \def\url#1{{\tt #1}}\fi
\expandafter\ifx\csname urlprefix\endcsname\relax\def\urlprefix{URL }\fi
\providecommand{\eprint}[2][]{\url{#2}}

\bibitem{feynman1977feynman}
Feynman R, Leighton R and Sands M 2010 {\em The Feynman Lectures on Physics\/}
  v. 3 (New York, USA: Basic books) ISBN 9780201021189

\bibitem{Popper1982-POPQTA}
Popper K~R 1982 {\em Quantum Theory and the Schism in Physics\/} (New York,
  USA: Routledge)

\bibitem{schrodinger}
Schr\"odinger E 1935 {\em Naturwissenschaften\/} {\bf 23} 807--812

\bibitem{ballentine1998quantum}
Ballentine L 1998 {\em Quantum Mechanics: A Modern Development\/} (Singapore:
  World Scientific) ISBN 9789810241056

\bibitem{2007arXiv0712.0149W}
{Wallace} D 2007 {\em arXiv e-prints\/} arXiv:0712.0149 (\textit{Preprint}
  \eprint{0712.0149})

\bibitem{lande4}
Land\'{e} A 1975 {\em American Journal of Physics\/} {\bf 43} 701

\bibitem{leblonde}
L\'{e}vy-Leblond J~M 1976 {\em American Journal of Physics\/} {\bf 44} 1130

\bibitem{PhysRevLett.122.053204}
Pursehouse J, Murray A~J, W\"atzel J and Berakdar J 2019 {\em Phys. Rev.
  Lett.\/} {\bf 122}(5) 053204
  \urlprefix\url{https://link.aps.org/doi/10.1103/PhysRevLett.122.053204}

\bibitem{duane}
Duane W 1923 {\em Proceedings of the National Academy of Sciences\/} {\bf 9}
  158--164 (\textit{Preprint}
  \eprint{https://www.pnas.org/doi/pdf/10.1073/pnas.9.5.158})
  \urlprefix\url{https://www.pnas.org/doi/abs/10.1073/pnas.9.5.158}

\bibitem{Batelaan_2000}
Batelaan H 2000 {\em Contemporary Physics\/} {\bf 41} 369–381 ISSN 1366-5812
  \urlprefix\url{http://dx.doi.org/10.1080/00107510010001220}

\bibitem{lande1965}
Landé A 1965 {\em American Journal of Physics\/} {\bf 33} 123--127 ISSN
  0002-9505 (\textit{Preprint}
  \eprint{https://pubs.aip.org/aapt/ajp/article-pdf/33/2/123/11943010/123\_1\_online.pdf})
  \urlprefix\url{https://doi.org/10.1119/1.1971264}

\bibitem{kay2024escape}
Kay A 2024 {\em Escape from Shadow Physics: The Quest to End the Dark Ages of
  Quantum Theory\/} (New York, USA: Basic Books) ISBN 9781541675773

\bibitem{bertet_complementarity_2001}
Bertet P, Osnaghi S, Rauschenbeutel A, Nogues G, Auffeves A, Brune M, Raimond
  J~M and Haroche S 2001 {\em Nature\/} {\bf 411} 166--170 ISSN 1476-4687
  publisher: Nature Publishing Group
  \urlprefix\url{https://www.nature.com/articles/35075517}

\bibitem{zeng2023}
Zeng T~H, Li K, Wang F, Shao B and Liang S~D 2023 {\em International Journal of
  Theoretical Physics\/} {\bf 62} 60
  https://www.researchsquare.com/article/rs-2243686/latest.pdf
  \urlprefix\url{https://app.dimensions.ai/details/publication/pub.1156068340}

\bibitem{Strnad_1985}
Strnad J and Kuhn W 1985 {\em European Journal of Physics\/} {\bf 6} 176
  \urlprefix\url{https://dx.doi.org/10.1088/0143-0807/6/3/009}

\bibitem{Pucci_Harris_Faria_Bush_2018}
Pucci G, Harris D~M, Faria L~M and Bush J~W~M 2018 {\em Journal of Fluid
  Mechanics\/} {\bf 835} 1136–1156

\bibitem{1933PCPS...29..297K}
{Kapitza} P~L and {Dirac} P~A~M 1933 {\em Proceedings of the Cambridge
  Philosophical Society\/} {\bf 29} 297

\bibitem{1986PhRvL..56..827G}
{Gould} P~L, {Ruff} G~A and {Pritchard} D~E 1986 {\em Phys. Rev. Lett.\/} {\bf
  56} 827--830

\bibitem{2001Natur.413..142F}
{Freimund} D~L, {Aflatooni} K and {Batelaan} H 2001 {\em Nature\/} {\bf 413}
  142--143

\bibitem{breit1923}
Breit G 1923 {\em Proc. Natl. Acad. Sci.\/} {\bf 9} 238–243

\bibitem{VANVLIET196797}
{Van Vliet} K 1967 {\em Physica\/} {\bf 35} 97--106 ISSN 0031-8914
  \urlprefix\url{https://www.sciencedirect.com/science/article/pii/0031891467901383}

\bibitem{VANVLIET20101585}
{Van Vliet} C~M 2010 {\em Physica A: Statistical Mechanics and its
  Applications\/} {\bf 389} 1585--1593 ISSN 0378-4371
  \urlprefix\url{https://www.sciencedirect.com/science/article/pii/S0378437109010401}

\bibitem{WENNERSTROM2014105}
Wennerström H 2014 {\em Advances in Colloid and Interface Science\/} {\bf 205}
  105--112 ISSN 0001-8686 special Issue in honor of Bjorn Lindman
  \urlprefix\url{https://www.sciencedirect.com/science/article/pii/S0001868613001401}

\bibitem{2018OptEn..57a5105M}
{Mobley} M~J 2018 {\em Optical Engineering\/} {\bf 57} 015105

\bibitem{bloch1929}
Bloch F 1929 {\em Zeitschrift f\"ur physik\/} {\bf 52} 555--600

\bibitem{mossbauer}
M\"{o}ssbauer R 1958 {\em Z. Physik\/} {\bf 151} 124--143

\bibitem{mossbauer2}
Eyges L 1965 {\em American Journal of Physics\/} {\bf 33} 790--802 ISSN
  0002-9505 (\textit{Preprint}
  \eprint{https://pubs.aip.org/aapt/ajp/article-pdf/33/10/790/12051300/790\_1\_online.pdf})
  \urlprefix\url{https://doi.org/10.1119/1.1970986}

\bibitem{mossbauer3}
Vandegrift G and Fultz B 1998 {\em American Journal of Physics\/} {\bf 66}
  593--596 ISSN 0002-9505 (\textit{Preprint}
  \eprint{https://pubs.aip.org/aapt/ajp/article-pdf/66/7/593/12210319/593\_1\_online.pdf})
  \urlprefix\url{https://doi.org/10.1119/1.18911}

\bibitem{tamvakis2019}
Tamvakis K 2019 {\em Basic quantum mechanics\/} (Switzerland: Springer Nature)

\bibitem{townsend2012modern}
Townsend J 2012 {\em A Modern Approach to Quantum Mechanics\/} (California,
  USA: University Science Books) ISBN 9781891389788

\bibitem{mcintyre2012quantum}
McIntyre D, Manogue C and Tate J 2012 {\em Quantum Mechanics\/} (London, UK:
  Pearson Education) ISBN 9780321850003

\bibitem{rae2002}
Rae A 2002 {\em Quantum mechanics\/} 4th ed (Bristol, UK: Institute of Physics)

\bibitem{Sakurai:1167961}
Sakurai J~J 1985 {\em Modern quantum mechanics\/} (Reading, MA: Addison-Wesley)

\bibitem{Griffiths2004Introduction}
Griffiths D 2005 {\em Introduction to Quantum Mechanics (2nd International
  Edition)\/} (New Jersey, USA: Pearson Prentice Hall)

\bibitem{moore1998}
Moore T~A 2017 {\em Six ideas that shaped physics: particles behave like
  waves\/} (New York, USA: McGraw-Hill Education)

\bibitem{kennard_zur_1927}
Kennard E~H 1927 {\em Zeitschrift für Physik\/} {\bf 44} 326--352 ISSN
  0044-3328 \urlprefix\url{https://doi.org/10.1007/BF01391200}

\bibitem{cowley-diffraction}
Cowley J~M 1995 {\em Diffraction Physics\/} 3rd ed (Amsterdam, Netherlands:
  Elsevier)

\bibitem{electron-phonon1}
Chase T, Trigo M, Reid A~H, Li R, Vecchione T, Shen X, Weathersby S, Coffee R,
  Hartmann N, Reis D~A, Wang X~J and Dürr H~A 2016 {\em Applied Physics
  Letters\/} {\bf 108} 041909 ISSN 0003-6951 (\textit{Preprint}
  \eprint{https://pubs.aip.org/aip/apl/article-pdf/doi/10.1063/1.4940981/14475001/041909\_1\_online.pdf})
  \urlprefix\url{https://doi.org/10.1063/1.4940981}

\bibitem{electron-phonon2}
Kurtz F, Dauwe T, Yalunin S, Storeck G, Horstmann J~G, B\"ockmann H and Ropers
  C 2024 {\em Nat. Mater.\/} {\bf 23} 890--897

\bibitem{photon-phonon}
Trigo M, Chen J, Vishwanath V~H, Sheu Y~M, Graber T, Henning R and Reis D~A
  2010 {\em Phys. Rev. B\/} {\bf 82}(23) 235205
  \urlprefix\url{https://link.aps.org/doi/10.1103/PhysRevB.82.235205}

\bibitem{grisenti_determination_1999}
Grisenti R~E, Schöllkopf W, Toennies J~P, Hegerfeldt G~C and Köhler T 1999
  {\em Physical Review Letters\/} {\bf 83} 1755--1758 publisher: American
  Physical Society
  \urlprefix\url{https://link.aps.org/doi/10.1103/PhysRevLett.83.1755}

\bibitem{Passon2022}
{Mairhofer} L and {Passon} O 2022 {\em Foundations of Physics\/} {\bf 52} 32
  (\textit{Preprint} \eprint{2203.08691})

\bibitem{Batelaan_2016}
Batelaan H, Jones E, Huang W~C~W and Bach R 2016 {\em Journal of Physics:
  Conference Series\/} {\bf 701} 012007 ISSN 1742-6596
  \urlprefix\url{http://dx.doi.org/10.1088/1742-6596/701/1/012007}

\bibitem{bernstein_measurement_1987}
Bernstein H~J and Low F~E 1987 {\em Physical Review Letters\/} {\bf 59}
  951--953 publisher: American Physical Society
  \urlprefix\url{https://link.aps.org/doi/10.1103/PhysRevLett.59.951}

\bibitem{klein_longitudinal_1983}
Klein A~G, Opat G~I and Hamilton W~A 1983 {\em Physical Review Letters\/} {\bf
  50} 563--565 publisher: American Physical Society
  \urlprefix\url{https://link.aps.org/doi/10.1103/PhysRevLett.50.563}

\bibitem{kaiser_direct_1983}
Kaiser H, Werner S~A and George E~A 1983 {\em Physical Review Letters\/} {\bf
  50} 560--563 publisher: American Physical Society
  \urlprefix\url{https://link.aps.org/doi/10.1103/PhysRevLett.50.560}

\bibitem{jabs_comment_1987}
Jabs A and Ramos R 1987 {\em Physical Review Letters\/} {\bf 58} 2274--2274
  publisher: American Physical Society
  \urlprefix\url{https://link.aps.org/doi/10.1103/PhysRevLett.58.2274}

\bibitem{liu_einsteinbohr_2015}
Liu X~J, Miao Q, Gel'mukhanov F, Patanen M, Travnikova O, Nicolas C, Ågren H,
  Ueda K and Miron C 2015 {\em Nature Photonics\/} {\bf 9} 120--125 ISSN
  1749-4893 publisher: Nature Publishing Group
  \urlprefix\url{https://www.nature.com/articles/nphoton.2014.289}

\bibitem{kopf_conservation_2025}
Kopf L, Barros R, Prabhakar S, Giese E and Fickler R 2025 {\em Physical Review
  Letters\/} {\bf 134} 203601 publisher: American Physical Society
  \urlprefix\url{https://link.aps.org/doi/10.1103/PhysRevLett.134.203601}

\bibitem{aharonov_conservation_nodate}
Aharonov Y, Popescu S and Rohrlich D 2025 {\em Proceedings of the National
  Academy of Sciences of the United States of America\/} {\bf 120} e2220810120
  ISSN 0027-8424
  \urlprefix\url{https://pmc.ncbi.nlm.nih.gov/articles/PMC10576110/}

\bibitem{collins_conservation_2025}
Collins D and Popescu S 2025 {\em Quantum\/} {\bf 9} 1815 publisher: Verein zur
  Förderung des Open Access Publizierens in den Quantenwissenschaften
  \urlprefix\url{https://quantum-journal.org/papers/q-2025-07-29-1815/}

\bibitem{Jack_2008}
Jack B, Padgett M~J and Franke-Arnold S 2008 {\em New Journal of Physics\/}
  {\bf 10} 103013
  \urlprefix\url{https://dx.doi.org/10.1088/1367-2630/10/10/103013}

\bibitem{Sroor_2021}
Sroor H, Moodley C, Rodríguez-Fajardo V, Zhan Q and Forbes A 2021 {\em Journal
  of the Optical Society of America A\/} {\bf 38} 1443 ISSN 1520-8532
  \urlprefix\url{http://dx.doi.org/10.1364/JOSAA.432431}

\bibitem{greenberger_simultaneous_1988}
Greenberger D~M and Yasin A 1988 {\em Physics Letters A\/} {\bf 128} 391--394
  ISSN 0375-9601
  \urlprefix\url{https://www.sciencedirect.com/science/article/pii/0375960188901144}

\bibitem{PhysRevA.65.033818}
Walborn S~P, Terra~Cunha M~O, P\'adua S and Monken C~H 2002 {\em Phys. Rev.
  A\/} {\bf 65}(3) 033818
  \urlprefix\url{https://link.aps.org/doi/10.1103/PhysRevA.65.033818}

\end{thebibliography}

\end{document}